\documentstyle[epsf,aps,prl]{revtex}
\draft
\begin{document}
\twocolumn[\hsize\textwidth\columnwidth\hsize\csname
@twocolumnfalse\endcsname
\draft\pagenumbering{roma}
\author{Liu Yu-Xi and Sun Chang-Pu}
\address{ Institute of Theoretical Physics, Academia Sinica, P.O.Box 2735,
Beijing 100080, China}
\title{Hybrid exciton-polaritons  in a bad microcavity containing  the organic 
and inorganic quantum wells}
\maketitle
\thispagestyle{empty}
\vspace{15mm}
\begin{abstract}
We study the hybrid exciton-polaritons in a bad microcavity containing the 
organic and inorganic quantum wells.  The corresponding polariton states 
are given. The  analytical solution and the numerical result of the  
stationary  spectrum for the cavity field are finished.
\end{abstract}
\vspace{4mm}
\vspace{3.cm}
\vskip2pc]
\newpage
\pagenumbering{arabic}

\begin{center}{\bf 1. Introduction} \end{center} 

Quantum wells (QWs) embedded in semiconductor microcavity structures have been
the subject of extensive theoretical and experimental investigation. We know
that the excitons play a fundamental role in the optical properties of the QWs.
The photodevices of excitons  may have small size, low power 
dissipation, rapidness and high efficiency.  All of above these are required
to the integrated photoelectric circuits.

The excitons are classified as Wannier excitons (they have large radius and
weak oscillator strength) and Frenkel excitons (they have  smaller radius and 
strong oscillator strength) by the size of the exciton model radius.     
  
Recently, a new excitionic state--hybrid exciton state in the composite
organic and inorganic  semiconductor heterostructure has been described by
pioneering work~\cite{A,B,C,D,E,F,G}. Since then from quantum well (QW) to 
quantum dot, the hybrid exciton states  due to resonant 
mixing of Frenkel and Wannier-Mott excitons have been demonstrate. The 
reference~\cite{A} shows that the hybrid excitons possess a strong oscillator strength and a 
small saturation density (or large radius). The reference~\cite{C} proposes to 
couple Frenkel excitons and Wannier-Mott excitons through a ideal microcavity.

There are two different coupling regime into which interaction between the 
cavity field and the optical transition of the excitons can be classified.
One is the weak coupling regime which the exciton-photon coupling is very
small and can be treated as a perturbation to the eigenstates of the uncoupled
exciton-photon system. Another is the strong coupling regime which the
exciton-photon coupling is so strong that it no longer be treated as 
perturbation. In the strong coupling regime, the QWs excitons emit photons
into the cavity. The photon are bounced back  by the mirror and reabsorbed by 
the QWs to create excitons again. So the Rabi oscillation are formed. The
exciton-polaritons mode splitting in a semiconductor microcavity is also
observed by many experimental group, such as~\cite{K,L}

In this paper, we will deal with the hybrid exciton-polaritons states for the
organic and in organic QWs in a bad cavity. In the section 2, we use
the  motion equations included damping effect to give the mixed hybrid-exciton
and cavity field modes,	that is, hybrid exciton-polaritons. In the section 3,
we will give the emission spectrum of the system in the case of the 
stationary state and the corresponding numerical results also are given.
In section 4, a simple conclusion is given.  

\begin{center}
{\bf 2. Model and exciton-polariton states }  
\end{center}
We begin with the Hamiltonian of the organic and inorganic quantum wells in
a ideal microcavity~\cite{C}

\begin{eqnarray}
H &=&\sum_{k}[\hbar\omega_{W}A^{+}_{k}A_{k}+\hbar\omega_{F}B^{+}_{k}B_{k}
+\hbar\Omega a^{+}_{k}a_{k}] \nonumber \\ 
&+&\sum_{k}[\hbar\Gamma_{13}(A^{+}_{k}a_{k}+A_{k}a^{+}_{k})+
\hbar\Gamma_{23}(B^{+}_{k}a_{k}+B_{k}a^{+}_{k})].
\end{eqnarray}
where $A_{k}$($A^{+}_{k}$), $B_{k}$($B^{+}_{k}$) and $a_{k}$($a^{+}_{k}$) 
are usual boson operators for Wannier, Frenkel and cavity field.
$\Gamma_{13}=\frac{1}{\hbar}P^{01}_{W}E_{0}$ and 
$\Gamma_{23}=\frac{1}{\hbar}P^{01}_{F}E_{0}$, $P^{01}_{W,F}$ is the moment 
matrix element for Wannier and Frenkel exciton from the ground state. 
$E_{0}$ is the amplitude of the vacuum electric field at the center of the 
cavity. This Hamiltonian is linear. For a bad microcavity, we could solve it 
by the motion equation included the damping coefficient and obtain  any mixed 
solutions of the hybrid exciton and cavity field.
 
But here, we only deal with the case of a single  mode cavity field. That is, 
the above equation is simplified into:  
\begin{eqnarray}
H &=&\hbar\omega_{W}A^{+}A+\hbar\omega_{F}B^{+}B
+\hbar\Omega a^{+}a \nonumber \\ 
&+&\hbar\Gamma_{13}(A^{+}a+Aa^{+})+
\hbar\Gamma_{23}(B^{+}a+Ba^{+}).
\end{eqnarray}
So we have the motion equation
\begin{mathletters}
\begin{eqnarray}
\frac{\partial a}{\partial t}&=&-i\Omega a-i\Gamma_{13}A-i\Gamma_{23}B
-\gamma_{1} a \\
\frac{\partial A}{\partial t}&=&-i\omega_{W}A-i\Gamma_{13}a-\gamma_{2}A \\
\frac{\partial B}{\partial t}&=&-i\omega_{F}B-i\Gamma_{23}a-\gamma_{3}B
\end{eqnarray}
\end{mathletters} 
In order to describe the properties of the bad cavity, The damping
coefficient  $\gamma_{i}$  are  added  phenomenologically to 
the above equation (3.a-3.c). In fact, when we write out the interaction between
the system and the reservoir, we could give a motion equation which includes
 the fluctuation terms and dissipative terms by  the Markov approximation. 
However, because we want to discuss the exciton-polaritons in the strong 
coupling regime. We aren't interested in the noise properties of the system.
So the fluctuation terms may be neglected. 

By use of the  Fourier transformation, we have:
\begin{mathletters}
\begin{eqnarray}
&&(i\gamma_{1}+\omega-\Omega)a(\omega)=ia(0)+\Gamma_{13}A(\omega)
+\Gamma_{23}B(\omega) \\
&&(i\gamma_{2}+\omega-\omega_{W})A(\omega)=iA(0)+\Gamma_{13}a(\omega)\\
&&(i\gamma_{3}+\omega-\omega_{F})B(\omega)=iB(0)+\Gamma_{23}a(\omega)
\end{eqnarray}
\end{mathletters} 
Where, $a(0)$, $A(0)$ and $B(0)$ are initial operators for the cavity field, 
Wannier excitons and Frenkel excitons respectively. 
In order to obtain $a(t)$, we need solve the pole equation
\begin{eqnarray}
&&(i\gamma_{1}+\omega-\Omega)(i\gamma_{2}+\omega-\omega_{W})
(i\gamma_{3}+\omega-\omega_{F})-\nonumber \\
&&(i\gamma_{2}+\omega-\omega_{W})\Gamma^{2}_{23}
-(i\gamma_{3}+\omega-\omega_{F})\Gamma^{2}_{13}=0
\end{eqnarray}
The solutions of this cubic equation could  be obtained analytically by
using any mathematics handbook, such as ~\cite{J}. But usually, a cubic
equation can be solved more quickly with numerical methods than with 
analytical procedures. So, we set the form of the analytical solutions of
 the equation (5) are $\omega_{1}=\omega^{\prime}_{1}-i\Gamma_{1}$, 
$\omega_{2}=\omega^{\prime}_{2}-i\Gamma_{2}$ and $\omega_{2}=\omega^{\prime}_{3}-i\Gamma_{3}$ respectively. 
If we make 
\begin{eqnarray}
&&F(\omega)=i(i\gamma_{2}+\omega-\omega_{W})
(i\gamma_{3}+\omega-\omega_{F})a(0) + \nonumber \\
&&+i\Gamma_{23}(i\gamma_{2}+\omega-\omega_{W})B(0)+
i\Gamma_{13}(i\gamma_{3}+\omega-\omega_{F})A(0)
\end{eqnarray}
We have $a(t)$ as following:
\begin{equation}
a(t)=\frac{F(\omega_{1})}{\Delta_{1}\Delta_{2}} e^{-i\omega_{1}t}
-\frac{F(\omega_{2})}{\Delta_{2}\Delta_{3}} e^{-i\omega_{2}t}
+\frac{F(\omega_{3})}{\Delta_{1}\Delta_{3}} e^{-i\omega_{3}t}
\end{equation}
with $\Delta_{1}=\omega_{1}-\omega_{2}$, $\Delta_{2}=\omega_{1}-\omega_{3}$
and $\Delta_{3}=\omega_{2}-\omega_{3}$. $\omega_{i}$ are determined by the 
pole equation. This equation indicates that the strong coupling of the
two kinds of the QWs exciton states and cavity field results in three new 
eigenstates.  Their eigenvalues are $\omega_{1}$, $\omega_{2}$ and $\omega_{3}$ 
respectively. These states are just hybrid exciton-polaritons states.  Their 
energy splitting 
are $\Delta_{1}$, $\Delta_{2}$ and $\Delta_{3}$ respectively

\begin{center}
{\bf  3. Stationary spectrum}
\end{center} 
For the case  of the ergodic and stationary process, the emission spectrum of 
the system is defined as following~\cite{H}:
\begin{equation}
S(\omega)=\int_{0}^{\infty}e^{-i\omega t}<a^{+}(t)a(0)>{\rm d}t + c.c.
\end{equation}
If  the cavity field, Wannier exciton and Frenkel exciton
are  initially  in  the number states $|n_{c}>$, $|n_{W}>$  and $|n_{F}>$ 
respectively, then 

\begin{equation}
<a^{+}(t)a(0)>=\bar{n}_{c}[-i\frac{E(\omega_{1})}{\Delta^{*}_{1}\Delta^{*}_{2}}
e^{i\omega_{1}t}+
i\frac{E(\omega_{2})}{\Delta^{*}_{2}\Delta^{*}_{3}}
e^{i\omega_{2}t}
-i\frac{E(\omega_{3})}{\Delta^{*}_{1}\Delta^{*}_{3}}
e^{i\omega_{3}t}]
\end{equation}

where $\bar{n}_{c}$ is mean photon number of the cavity field and 
\begin{equation}
E(\omega)=(i\gamma_{2}+\omega-\omega_{W})
(i\gamma_{3}+\omega-\omega_{F})
\end{equation}
As the general exciton-polaritons~\cite{M}, If we assume that the damping is 
moderate, the process is almost ergodic and stationary. It's deserved 
to point out that all of the parameters excepting for $\omega$ are fixed by the 
properties of the organic and inorganic QWs as well as microcavity material.
We always may choose some moderate parameters so that the stationary condition
could be satisfied.
So the spectrum of the system is :

\begin{equation}
S(\omega)=\frac{A}{(\omega-\omega^{\prime}_{1})]^{2}+\Gamma^{2}_{1}}
+\frac{B}{(\omega-\omega^{\prime}_{2})]^{2}+\Gamma^{2}_{2}}
+\frac{C}{(\omega-\omega^{\prime}_{3})]^{2}+\Gamma^{3}_{2}}
\end{equation}
with
\begin{mathletters}
\begin{eqnarray}
A(\omega)&=&2\bar{n}_{c}\frac{Re [E(\omega_{1})\Delta_{1}\Delta_{2}(\omega-\omega_{1})]}
{|\Delta_{1}|^{2}|\Delta_{2}|^{2}} \\
B(\omega)&=&2\bar{n}_{c}\frac{Re [E(\omega_{2})\Delta_{3}\Delta_{2}(\omega-\omega_{2})]}
{|\Delta_{3}|^{2}|\Delta_{2}|^{2}} \\
C(\omega)&=&2\bar{n}_{c}\frac{Re [E(\omega_{3})\Delta_{3}\Delta_{1}(\omega-\omega_{3})]}
{|\Delta_{3}|^{2}|\Delta_{1}|^{2}}
\end{eqnarray} 
\end{mathletters}
We find that when the system reaches stability, the hybrid exciton-polaritons
 spectrum is superposition three Lorentzian lines which are expected. The 
 exciton-polariton splitting may be 
measured at the peak points of the emission spectrum which are determined by
the condition $\frac{d S(\omega)}{d \omega}=0$.

We apply the general  eq.(11) to give a  numerical sketch map. 
We firstly adopt to the assumption of the reference~\cite{C} namely 
$\omega_{F}=\Omega$, $\omega_{W}=\omega_{F}(1+\delta)$, $\delta=10^{-2}$.

 Now we give a set of the possible values for the above parameters.
 $\bar{n}_{c}$ only determines the amplitude of the
 spectrum, so we set $\bar{n}_{c}=1$.
  We assume that 
 $\omega_{F}=\Omega=1562 meV$, $\Gamma^{2}_{23}=16meV$, $\Gamma^{2}_{13}=8meV$, 
 $\gamma_{1}=0.1meV$, $\gamma_{2}=0.18meV$, $\gamma_{3}=0.12meV$. By using
 these numbers and eq.(11), we give the sketch map of the  spectrum for the
 system. (Fig.1). 
 
\begin{figure}
\epsfxsize=6 cm
\centerline{\epsffile{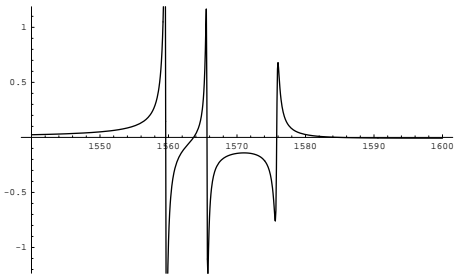}}
\caption[]{Schematic drawing for the emission spectrum }
\end{figure}

 This sketch map shows there are sudden change near the three peaks. 
  If we choose moderate parameters, 
 $A(\omega)$, $B(\omega)$ and $C(\omega)$ are 
slowing varing functions of $\omega$ near the peaks and can be considered as
 constant. So the sudden varing points will disappear,
  the precise stationary spectrum is given.

\begin{center}
{\bf 4. Conclusion}
\end{center}

In conclusion, the hybrid exciton-polariton states in a bad microcavity
 containing  the organic and inorganic quantum wells is given. Although 
 we only discuss a single mode model, this approach also could be apply
 to general case.  This paper shows that the hybrid exciton-polaritons
 decay at three difference rate. The analytical and numerical results of
 the emission spectrum for the exciton-polaritons are also given.

\end{document}